# An Analogue Front-End ASIC Prototype Designed For PMT Signal Readout

LIU Jian-Feng(刘建峰)[1,2]   ZHAO Lei(赵雷)[1,2; 1)]   YU Li(于莉)[1,2]   LIANG Yu(梁宇)[1,2]   QIN Jia-Jun(秦家军)[1,2]
YANG Yun-Fan(杨云帆)[1,2]   WU Wei-Hao(邬维浩)[1,2]   LIU Shu-Bin(刘树彬)[1,2]   AN Qi(安琪) [1,2]

1 State Key Laboratory of Particle Detection and Electronics, University of Science and Technology of China, Hefei, 230026, China,

2 Department of Modern Physics, University of Science and Technology of China, Hefei, 230026, China

**Abstract:** The Large High Altitude Air Shower Observatory (LHAASO) is designed for high energy gamma ray and cosmic ray detection. A Water Cherenkov Detector Array which is sensitive to gamma ray showers above a few hundred GeV is proposed to survey gamma ray sources. The WCDA consists of 3600 PhotoMultiplier Tubes (PMT) which collect the Cherenkov light produced by the shower particles in water. Both high precision time and charge measurement are required over a large dynamic range from 1 photo electron (P.E.) to 4000 P.E. Prototype of an analogue front-end Application Specific Integrated Circuit (ASIC) fabricated in Chartered 0.35 μm CMOS technology is designed to read out PMT signal in the WCDA. This ASIC employs leading edge discrimination and $RC^4$ shaping structure; combined with the following Time-to-Digital Converter (TDC) and Analog-to-Digital Converter (ADC), both the arrival time and charge of the PMT signal can be measured. Initial test results indicate that time resolution is better than 350 ps and charge resolution is better than 10% at 1 P.E. and better than 1% with large input signals (300 P.E. to 4000 P.E.). Besides, this ASIC has a small channel-to-channel crosstalk and low ambient temperature dependency.

## 1 Introduction

The Large High Altitude Air Shower Observatory (LHAASO) is a multipurpose complexity which consists of different detectors for high energy gamma ray and cosmic ray detection [1]. A Water Cherenkov Detector Array (WCDA) in LHAASO will be built to survey the gamma ray sources at energies higher than 300 GeV. The WCDA is configured with four 150 m × 150 m water pools adjacent to each other. Each pool has 900 PMTs which face upward to observe the Cherenkov light produced in water by secondary particles induced by air showers [2]. The PMT signals are transmitted to the electronics via 30-m coaxial cables. The signal dynamic range of the PMT is from 1 P.E. to 4000 P.E. Within the large dynamic range, both precise time and charge measurement are required [3], as listed in Table 1.

Table 1. Measurement requirements of the WCDA readout electronics in LHAASO.

| Parameter | Requirement |
| --- | --- |
| Channel number | 3600 |
| Signal dynamic range | 1 P.E.~4000 P.E. |
| Resolution of charge measurement | 30%RMS@ 1 P.E. 3%RMS@ 4000 P.E. |
| RMS of time resolution | <0.5 ns RMS |

These requirements bring quite a few challenges in the readout electronics design. Firstly, the PMT signal varies from 1 P.E to 4000 P.E. and thus a large dynamic range signal processing capability of is required. Secondly, high time and charge measurement resolution within the full dynamic range is challenging, especially for 1 P.E signal, the peak current of which is as low as 60 μA (it will be further attenuated by the 30-m cable). This requires the readout electronics to have a sufficiently low noise level. Lastly, considering a 1-4000 dynamic range, the reflection of large signals would influence the measurement of subsequent small signals; therefore, high precision impedance matching is required.

*Supported by Knowledge Innovation Program of the Chinese Academy of Sciences (KJCX2-YW-N27), National Natural Science Foundation of China (11175174) and the CAS Center for Excellence in Particle Physics (CCEPP).
1) Email: zlei@ustc.edu.cn

Some large dynamic range readout ASICs for PMTs have been designed for other experiments [4-10]. However, the performance of these ASICs is not in full compliance with the requirement of WCDA in LHAASO.

In this paper, we present the design of a front-end ASIC for the time and charge measurement, to fulfill the requirement of LHAASO WCDA, with all the analogue circuits integrated within one chip. Details will be described in the following sections.

## 2 Circuits architecture

Fig. 1 shows the architecture of the PMT readout electronics in the WCDA of LHAASO. The signals from PMTs are sent to the readout electronics via 30-m coaxial cables, and manipulated by the ASIC. This ASIC generates two outputs: one is the output of the leading edge discrimination circuits and the other is Quasi-Gaussian signal produced by the shaping circuits, the amplitude of which corresponds to the charge information of the PMT signal. Combined with the Time-to-Digital Converter (TDC) integrated in the Field Programmable Gate Array (FPGA), time measurement can be achieved. An external Analog-to-Digital Converter (ADC) is used to digitize the Quasi-Gaussian signal; with the peak detection logic in the FPGA, the charge measurement can be achieved. Configuration of the ASIC, such as gain, test point selection and Digital-to-Analog Converter (DAC) input code, is also managed by the FPGA.

As for the ASIC, to cover a dynamic range up to 4000 P.E., we designed two readout channels for charge measurement: one processes the PMT anode signals to cover the range from 1 P.E. to 100 P.E.; the other is used for the signal readout of the PMT tenth dynode, to cover the range from 40 to 4000 P.E. The anode signal is also fed into leading edge discriminators for time measurement in the full scale range. To achieve high precision impedance matching, an external 50 Ω resistor is used for the input signal termination of the anode channel. As for the dynode channel, we use an external 20 dB π attenuation resistor network. It can achieve an equivalent input resistance of 50 Ω; meanwhile, by adjusting the dynode signal amplitude with this resistor network and the pre-amplifier, the anode and dynode can share a similar circuit structure. With this scheme, we designed the ASIC in Chartered 0.35 μm CMOS technology.

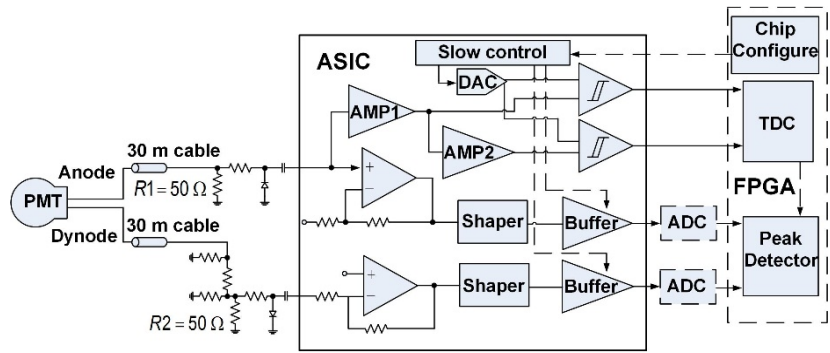

Fig. 1 Architecture of the PMT readout electronics in the WCDA of LHAASO.

### 2.1 Basic structure of the ASIC

The detailed diagram of the anode channel is shown in Fig. 2. The anode input current signal is converted to a voltage signal on the termination resistor. Then this voltage signal is split into two paths inside the ASIC. One path is for time measurement and the other is for charge measurement.

In the time measurement part, the signal is amplified by two amplifiers, and outputs of these two amplifies are fed into two discriminators, for which two different thresholds are selected. The thresholds are equivalent to 1/4 P.E. and 3 P.E. (user controlled in the test), respectively. The above design is aimed to avoid time resolution deterioration caused by noise or interference in the baseline of large input signals.

The pre-amplifier of time measurement part adopts a non-inverting amplification structure which features a high input impedance to avoid degrading the impedance matching precision, and it contains a differential input Operational Transconductance Amplifier (OTA), marked as A1 in Fig. 2.

As for the charge measurement part, the input signal is firstly amplified by the OTA A2 which has a smaller bandwidth than A1. Then the amplified signal is filtered by an $RC^4$ shaping circuit, and then driven output by a class-AB buffer which has a rail-to-rail driving capability. The shaper employs a four-order passive low-pass filter with three OTAs as the isolation stages; with this structure it is easier to achieve good stability compared with active filters, e.g., Sallen-Key filter [11].

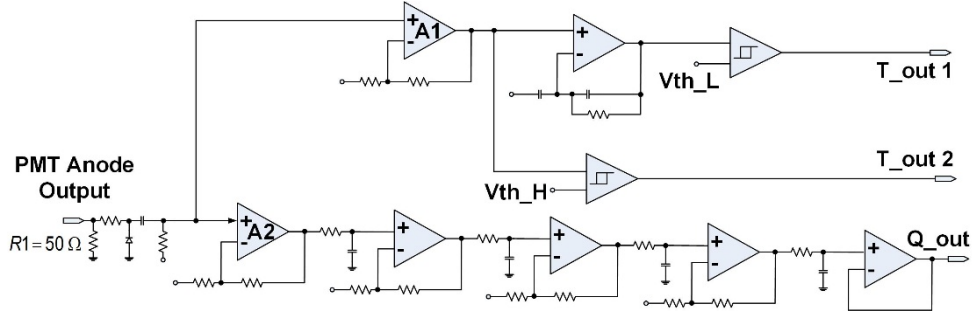

Fig. 2 Anode channel block diagram of the ASIC

As mentioned above, the structure of the dynode channel is quite similar to the charge measurement part of the anode channel. The difference is the pre-amplifier of dynode channel is an inverting amplifier.

### 2.2 Pre-amplifier design

Considering the high precision and large dynamic range requirements, the pre-amplifier should have the feature of low noise and large output swing. Since the noise and output swing of pre-amplifier is dictated by the OTA, the performance of the OTA should be optimized carefully.

The OTA A1 (in Fig. 2) employs two-stage amplifier with miller compensation structure (as shown in Fig. 3). The first stage incorporates differential input pair with active current mirror loads and the second is configured as common-source stage. As stated in the literature [12], the input equivalent noise density is given by the following formula ($\overline{v_{R1}^2}$, $\overline{v_{R2}^2}$ and $\overline{v_A^2}$ denote the power spectral density of resistor $R_1$, $R_2$ and OTA, respectively).

$$\overline{v_{ieq}^2} = \overline{v_{R1}^2}(\frac{R_2}{R_2+R_1})^2 + \overline{v_A^2} + \overline{v_{R2}^2}(\frac{R_1}{R_2+R_1})^2. \quad (1)$$

For a large close loop gain $\frac{R_2+R_1}{R_1}$, the Eq. (1) can be simplified to

$$\overline{v_{ieq}^2} = \overline{v_{R1}^2} + \overline{v_A^2}. \quad (2)$$

And the integrated input RMS noise can be calculated as

$$v_{n,in} = \sqrt{\int \overline{v_{ieq}^2} \frac{1}{1+(\frac{f}{f_{3dB}})^2} df}. \quad (3)$$

In Eq. (3) $f_{3dB}$ denotes the 3 dB bandwidth of the pre-amplifier, which is dictated by the input signal spectrum. And the 3 dB bandwidth of pre-amplifier is optimized to be 150 MHz within which 99% power of the PMT waveform is concentrated through FFT analysis. Therefore the Eq. (3) implies that small input RMS noise can be obtained by lowing the input RMS noise density. The noise density of OTA and $R_1$ can be approximately expressed as

$$\overline{v_A^2} = \frac{16}{3}kT\frac{1}{g_{m1}}(1+\frac{g_{m3}}{g_{m1}}), \quad (4)$$

$$\overline{v_{R1}^2} = 4kTR_1. \quad (5)$$

In Eq. (4) $g_{m1}$ and $g_{m3}$ denote the transconductance of input pair (M1, M2 in Fig. 3) and current mirror load (M3, M4 in Fig. 3) of OTA. Both increasing the input pair transconductance and decreasing resistance of $R_1$ can lowering the input noise, however increasing power consumption. Because the former is proportional to square root of drain current of MOSFET and the latter increases the driving current demand for a given close loop gain.

The input noise contribution of OTA is designed to approximate to the amount of $R_1$, which balances the total input noise density and power dissipation. Through simulation the transconductance of input pair and current mirror load are optimized to be about 6 mS and 2 mS, respectively, when the power dissipation is 3.6 mA Fig. 4 shows the simulated results for gain and equivalent input noise density of pre-amplifier. The simulated bandwidth is 146 MHz at the gain of 7 and total equivalent input noise is estimated to be 50 μV RMS (from DC to 1 GHz), which corresponds to 1/60 P.E., well beyond the measurement requirement even for 1 P.E. signal. Moreover, the OTA A1 has a high Power Supply Rejection Ratio (PSRR) of 62 dB, which is better than that of the single-ended input OTA often employed as the low noise pre-amplifiers [4-8].

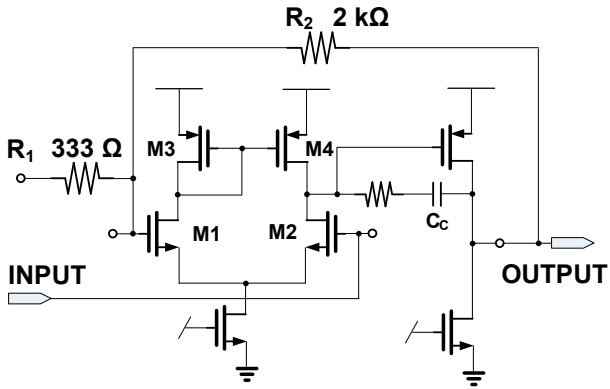

Fig. 3 OTA schematic in the pre-amplifier

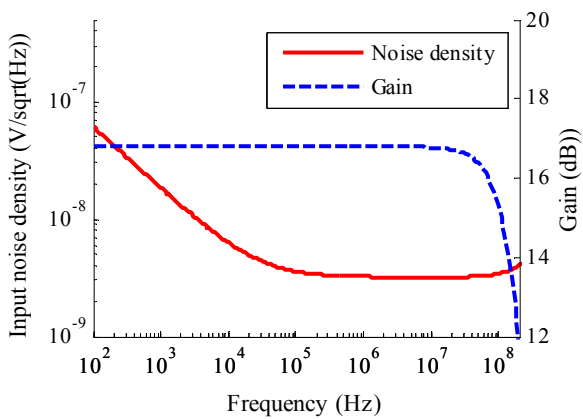

Fig. 4 Simulated gain and equivalent input noise density of pre-amp.

In addition to low noise requirement, the large output swing also demands the OTA of pre-amplifier is power hungry. Therefore the optimization of OTA are trade-offs between noise, power consumption, voltage headroom and bandwidth.

**2.3 Low temperature drift design**

The actual environmental temperature of the experiment will vary from 0 ºC to 50 ºC and temperature variation may influence the performance of the ASIC. Therefore a low temperature drift design for this ASIC is necessary. For charge measurement, the small signal transfer function of shaping circuit is a low-pass filter with gain about 40 dB, and both the gain and bandwidth may vary with temperature. By using the transfer function of the shaping circuit and inverse Laplace transform, the output peak value for $RC^n$ filter can be calculated as:

$$V_{o,M} = \frac{A_v}{\tau} \frac{(n-1)^{n-1}}{(n-1)!} e^{-(n-1)}, \quad (6)$$

where $A_v$ and $\tau$ is the gain and time constant of the shaping circuit, respectively. Furthermore, the bandwidth is inversely proportional to $\tau$, hence the peak value of output is proportional to the Gain Bandwidth Product (GBP).

The overall gain of shaping circuit equals to the product of close loop gain in each amplifier stage. And the close loop gain transfer function of single amplifier can be calculated as:

$$\frac{V_{out}}{V_{in}} = \frac{1}{\beta} \frac{1}{1+\frac{1}{A_v \beta}}. \quad (7)$$

Eq. (7) indicates that close loop gain is proportional to the reciprocal of feedback coefficient and is affected slightly by the open loop gain. The effect of open loop gain is negligible when it exceeds 60 dB and hence all amplifiers were designed with a large open loop gain. For feedback coefficients, they were chosen to employ the same type resistors which have the same temperature coefficient. As shown in Fig. 7, the simulation results indicate that the gain variation with the temperature is within 0.2%.

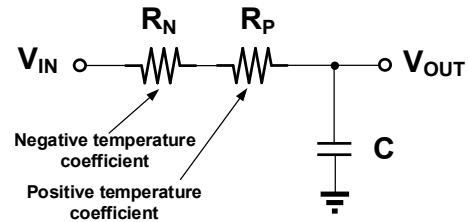

Fig. 5 Schematic of low temperature drift RC filter

Bandwidth of low pass filter is inversely proportional to the product of resistance and capacitance. The capacitor is almost temperature insensitive but resistor have a relatively large temperature coefficient Chartered 0.35 μm

CMOS technology. Hence we focus on the optimization of resistor. We can choose two type resistors (as shown in Fig. 5) having opposite sign temperature coefficients and combine them in series with an appropriate ratio to obtain a coefficient approach to zero. As shown in Fig. 6, the simulation results of the resistance variation indicates that the temperature coefficient of the combination resistor is much smaller than that of the original ones. By choosing the proper ratio of the two type resistors, as show in the Fig. 7, the simulation results show that the bandwitdth variation with temperature can achieve 0.6% in the range of 0 ºC to 50 ºC.

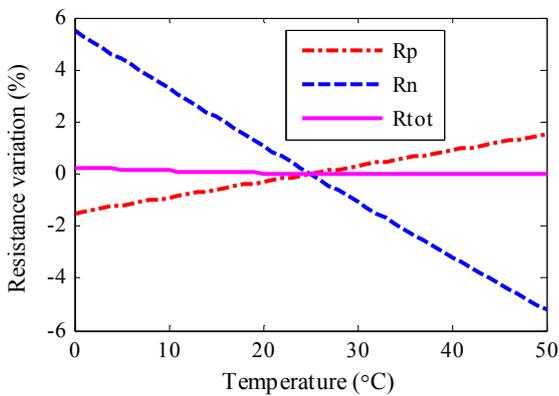

Fig. 6 Resistance variation with temperature

The gain and bandwidth are optimized to have negative and positive correlation with temperature, respectively, therefore the overall GBP changes very little with temperature (as shown in Fig. 7). According to Eq. (6) the temperature insensitive GBP implies that the shaping circuit output would have very weak dependency with temperature. The overall charge measurement temperature drift will be presented in the next section.

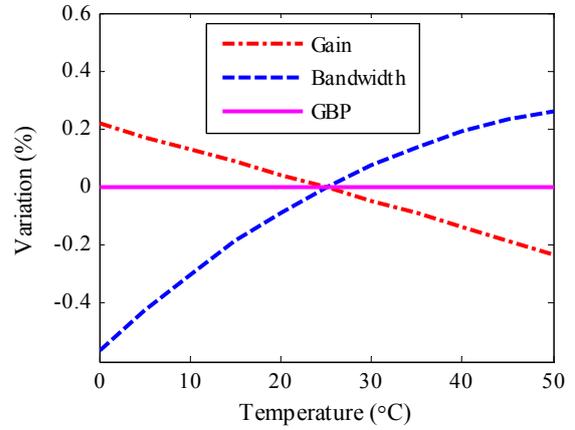

Fig. 7 Gain and bandwidth variation versus temperature.

**2.5 Output buffer**

To drive large swing shaping signal out of the chip, the shaping circuit demands an analog rail-to-rail output voltage buffer with high driving ability. The shaping signal will be driven to an external ADC through cable in the test. Hence the resistive load of output buffer is 50 Ω and parasitic capacitance is estimated to be about 10 pF. For an output peak reaching to 1.8 V, the output buffer needs to provide 36 mA dynamic current. To reduce the quiescent power dissipation of output buffer, the output buffer employs class-AB structure which is shown in Fig. 8 [13, 14].

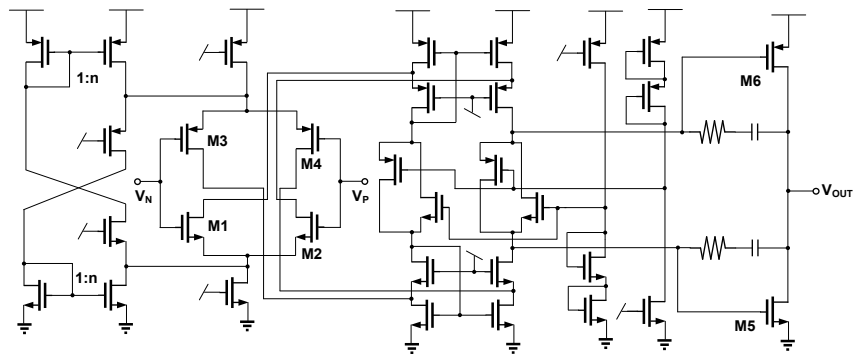

Fig. 8　Schematic of class-AB output buffer

The input stage incorporates complementary MOS transistors to achieve wide common mode input range. To maintain the gain, bandwidth and stability, the input transconductance should have a good flatness. To alleviate the transconductance variation in the transition of different type MOS transistors (M1, M2, M3 and M4 in Fig. 8), tail current compensation is incorporated. The relationship between proportion of compensation tail current and transconductance flatness is investigated. The simulation results shown in Fig. 9 indicate the transconductance variation is reduced to 10% with the ratio of 1:3 between the compensation tail current and normal tail current.

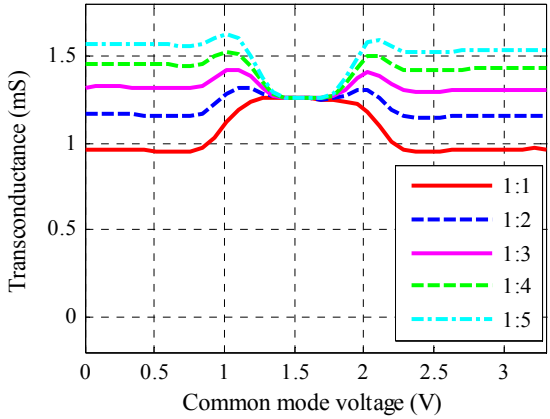

Fig. 9 Transconductance variation versus common mode voltage

To make efficient use of the supply voltage and supply current, the output stage of the output buffer in Fig. 8 incorporates class-AB biased output transistors. The class-AB action is performed by making both output transistors M5 and M6 (in Fig. 8) configuration as floating current source. The ratio of dynamic and quiescent current is about 16:1 and the output can provide sufficient current to drive off chip 50 Ω resistor in parallel with 10 pF capacitor in AC mode. As shown in Fig. 10, the output buffer can follow the input signal very well for a signal with 1.8 V peak voltage.

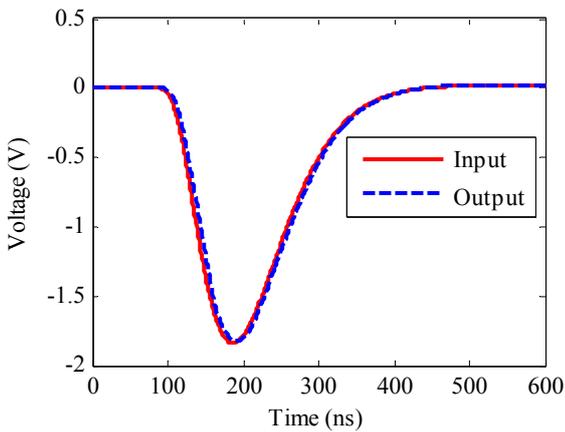

Fig. 10 Input and output waveform of class-AB buffer

## 3. Simulation

A series of simulations were conducted to evaluate the performance of this ASIC.

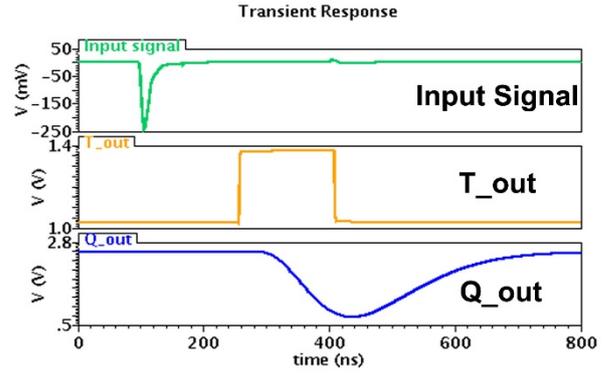

Fig. 11 Kernel nodes simulation waveform of the anode channel.

Firstly, to confirm the functionality of this ASIC, we conducted transient simulations. The simulation waveforms of the anode channel kernel nodes are shown in Fig. 11. The leading edge of discriminator output ("T_out" in Fig. 11) corresponds to the arrival time of PMT signals and the amplitude of shaper output ("Q_out" in Fig. 11) corresponds to the charge of PMT signals. The transient simulation indicates that this ASIC functions as expected.

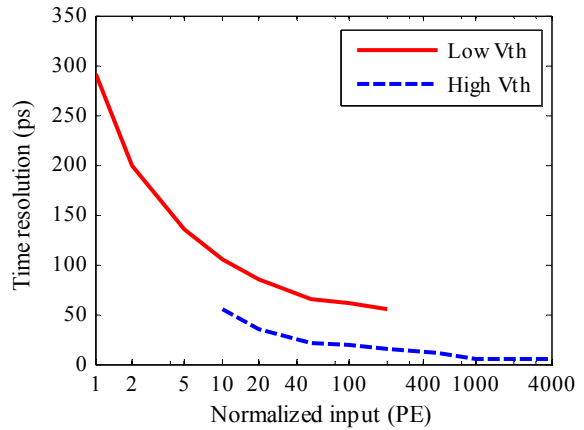

Fig. 12 Time resolution simulation results

Then, the time resolution evaluation was conducted by obtaining the voltage noise and the slope of input signal of discriminator through noise and transient simulation. As shown in Fig. 12, the time resolution of low threshold (1/4 P.E.) is estimated to be better than 300 ps and the high threshold (3 P.E.) is better than 50 ps. The time delay simulation results depicted in Fig. 13 show the time walk is less than 20 ns.

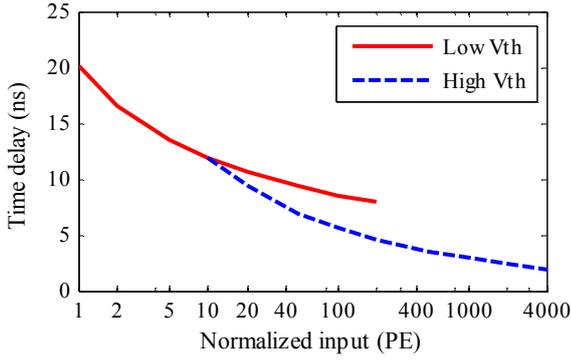

Fig. 13 Time walk simulation results.

As for charge measurement, the peak value of the shaper output corresponds to the input charge. We quantified the peak value with ideal a 12 bit 2 V input range ADC to obtain the charge transfer curve as shown in Fig. 14. By utilizing two readout channels, the charge measurement can cover 4000 dynamic range with a sufficient overlap. Moreover, by simulating the output noise of shaper, the simulated charge resolution can be obtained. As shown in Fig. 15 the simulation results are better than 5% in the full dynamic range.

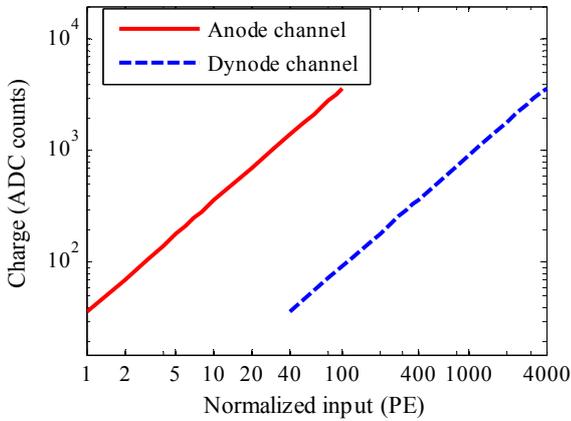

Fig. 14 Charge transfer curve simulation results

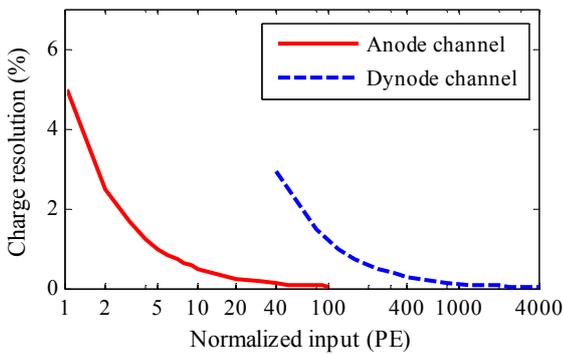

Fig. 15 Charge resolution simulation results

Finally, temperature drift simulations were conducted to verify the low drift design. The simulation results depicted in Fig. 16, Fig. 17 and Fig. 18 show the temperature dependency of time resolution, charge transfer curve and charge resolution, respectively. Within the temperature range of 0 ºC to 50 ºC, time resolution remains better than 350 ps, charge measurement results vary within ±1% and charge resolution remains better than 6%.

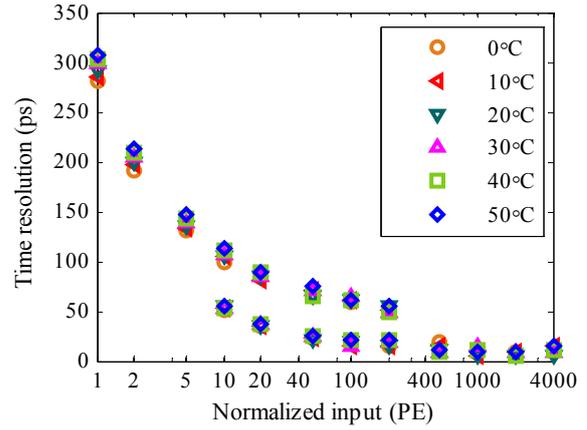

Fig. 16 Temperature dependency of time resolution

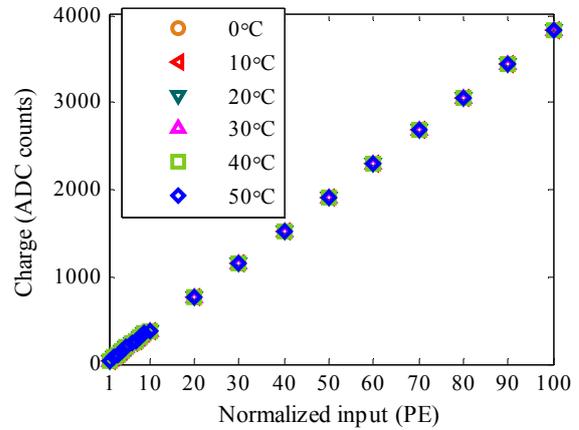

Fig. 17 Temperature dependency of charge transfer curve

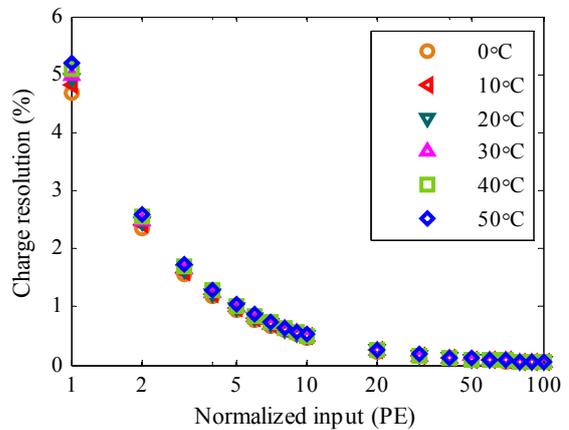

Fig. 18 Temperature dependency of charge resolution

After fabrication of the ASIC, we set up a test platform in the laboratory and conducted a series of tests, which will be presented in the next sections.

## 4. Circuits performance

### 4.1 Laboratory test set-up

A dedicated test board was designed to test this ASIC. The block diagram of test platform is constructed as shown in Fig. 19. We used an arbitrary waveform signal generator (Agilent Technologies 81160A) to generate the input signal for the test, according to the waveform of the PMT (R5912) output signal acquired by the oscilloscope (Lecroy 104 MXi).

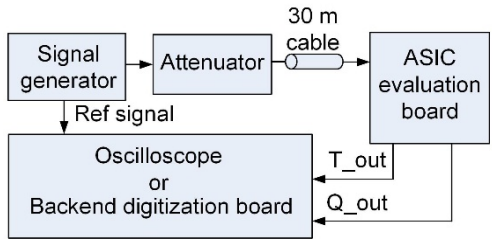

Fig. 19 Block diagram of the test platform.

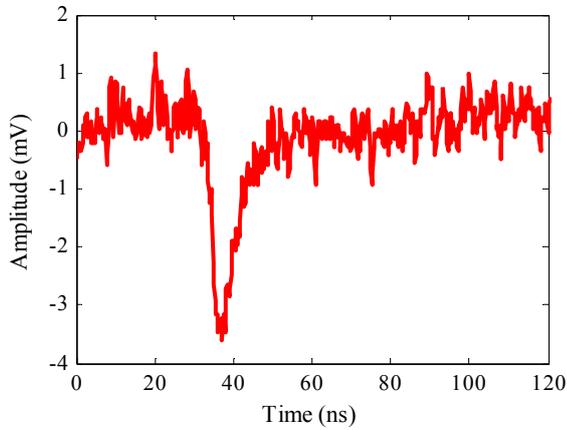

Fig. 20 Waveform of single P.E. PMT signal.

Fig. 20 shows the waveform of 1 P.E. PMT output signal, with a leading edge (10% to 90%) of 4.6 ns and a trailing edge (90% to 10%) of 16 ns. We used an attenuator (Wavetek Step Attenuator Model 5080.1) to adjust the input signal amplitude to estimate the ASIC performance in a large dynamic range. The output from the attenuator is transmitted to the ASIC evaluation board through a 30-m cable, to approximate the application situation. Then the time resolution can be tested using the oscilloscope, while the charge resolution test is conducted with a backend digitization board (with ADCs and an FPGA on it).

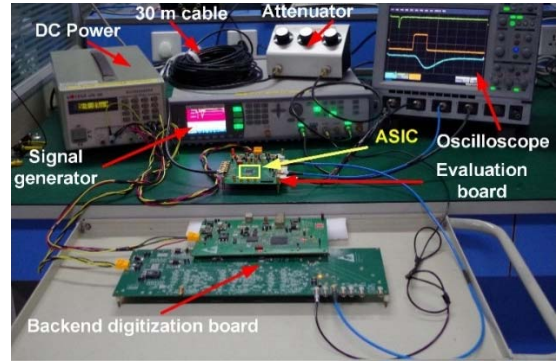

Fig. 21 Photograph of the test platform.

Fig. 21 shows the photograph of the test platform. The ASIC is marked in the middle with a yellow rectangle.

### 4.2 Basic functionality

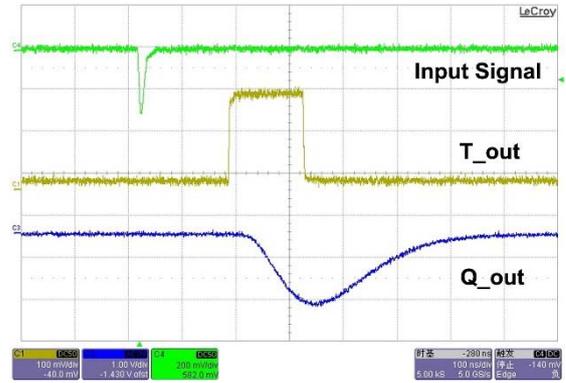

Fig. 22 Waveform test results.

Shown in Fig. 22 are waveform test results of the kernel nodes in the ASIC, which concord well with those in Fig. 11. The test results indicate that this prototype ASIC functions well.

### 4.3 Noise performance

The noise performance is investigated by measuring the output noise (RMS value) of three main blocks: pre-amplifier in time measurement part, shaper of anode channel, and shaper of dynode channel.

According to the test results, the intrinsic RMS noise of the oscilloscope (Tektronix DPO 7354C) is about 258.1 μV; by subtracting the above noise contribution, we can calculated the noise of the three output nodes. And then the worst case SNR is further calculated (with reference to 1 P.E. for anode channel and 40 P.E. for the dynode channel). The results are listed in Table 2, also compared with the simulation results. The test and simulation results concord well, which indicates that this ASIC has good noise performance, beyond the application requirement (30% charge resolution @ 1 P.E.).

Table 2. The noise performance of three main blocks.

|  | Pre-amplifier Test / Simulation | Anode shaper Test / Simulation | Dynode Shaper Test / Simulation |
| --- | --- | --- | --- |
| Output voltage | 18.73 mV / 18.33 mV | 17.6 mV / 17.9 mV | 17.84 mV / 18.4 mV |
| RMS noise | 0.76 mV / 0.60 mV | 1.12 mV / 0.84 mV | 0.84 mV / 0.52 mV |
| SNR | 24.64 / 30.55 | 15.71 / 21.28 | 21.23 / 35.38 |

### 4.4 Time measurement

As shown in Fig. 19, the signal source generates two synchronous signals: one is fed into the evaluation board through an attenuator as the input signal; the other is sent to the oscilloscope as a reference signal. The delay between the "T_out" in Fig. 19 and the reference signal is measured by an oscilloscope (Lecroy 104 MXi). The statistic mean and RMS value of the delay test results correspond to time walk and resolution, respectively (the jitter between input signal and reference signal is less than 10 ps, which is negligible).

The time walk and resolution test results are shown in Fig. 23 and Fig. 24. The two curves in each figure correspond to the outputs of the discriminators with a low (1/4 P.E.) and a higher (3 P.E., user controlled) threshold, respectively. The time walk is below 15 ns and it can be further calibrated according to the charge measurement results. The time resolution is better than 350 ps @ 1 P.E. and better than 100 ps with the large input signals, which concords well with the simulation results.

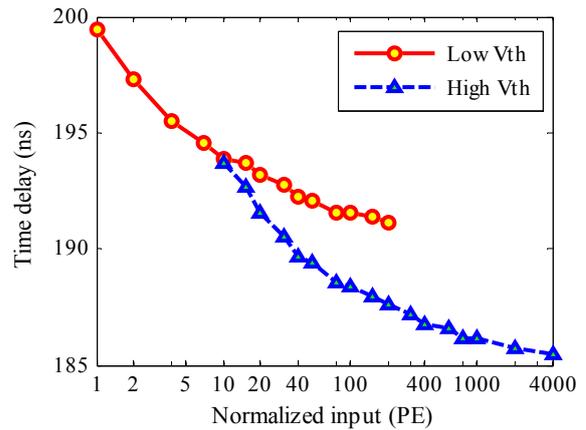

Fig. 24 Time walk test results.

### 4.5 Charge measurement

As shown in Fig. 19, the shaper output "Q_out" is imported to an ADC on the backend digitization board. With the logic of the FPGA device in the same board, the peak value of the ADC output signal can be obtained, which correspond to the charge measurement results.

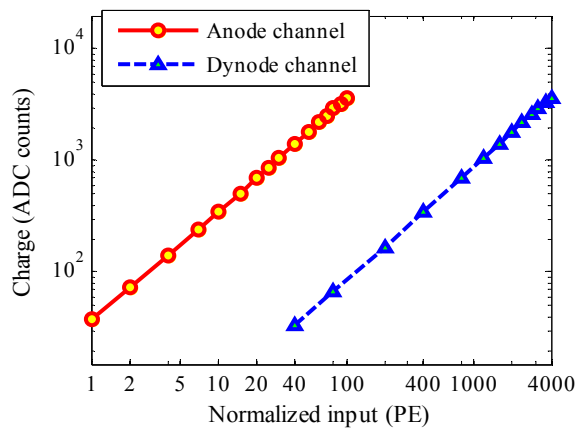

Fig. 25 Charge transfer curve test results

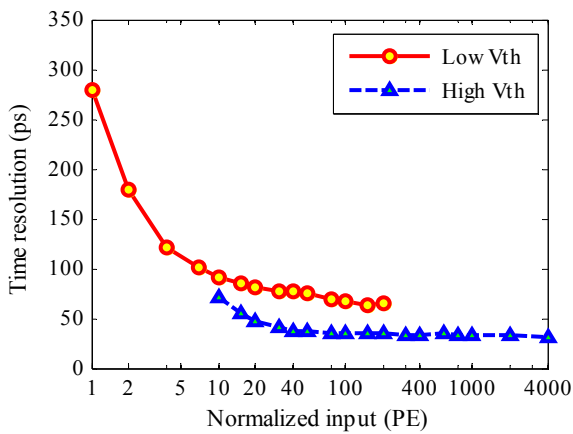

Fig. 23 Time resolution test results.

Fig. 25 shows the charge test results (with unit of ADC count) with different input signal amplitudes. The curves are approximately linear, with linear regression coefficients of R=0.9995 and R=0.9997 for anode and dynode channel, respectively. Each channel has a dynamic range about 100 and hence two readout channels can cover 4000 dynamic range with a sufficient overlap.

Fig. 26 shows the charge resolution in the full dynamic range. The overall charge resolution is better than 10 % and better than 1% with input signals larger than 300 P.E., which includes the effect of the ASIC noise and the input noise of the ADC. The input noise contribution of the ADC is about 1 mV, taking the ASIC shaper noise level (shown in Table 2) into account, and hence it has a greater impact on the dynode channel charge resolution.

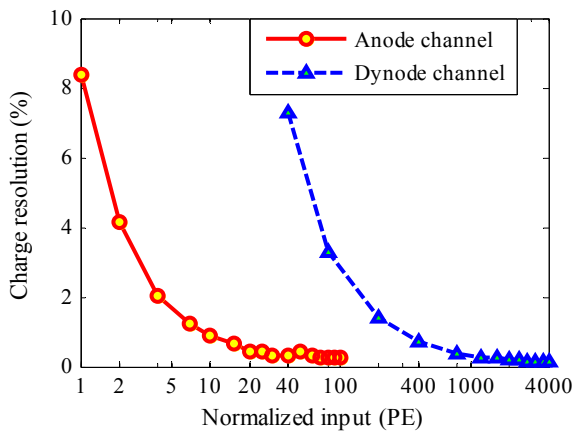

Fig. 26 Charge resolution test results

### 4.6 Crosstalk

Channel isolation can be evaluated by counting error hits caused by crosstalk due to neighboring channels. No error hits or charge performance deterioration was observed when large signals are imported to the neighboring channel. It indicates that this ASIC achieves a good channel-to-channel isolation performance.

### 4.7 Ambient temperature dependency

We also conducted temperature drift tests on this ASIC. The test temperature range is tuned from 0 ºC to 50 ºC which covers the temperature variation of the real experiment environment.

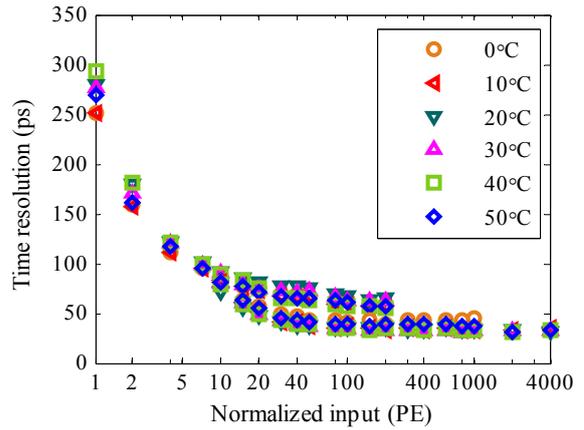

Fig. 27 Temperature dependency of time resolution

Fig. 27 shows the ambient temperature dependency of time measurement. Within the temperature range of 0 ºC to 50 ºC, time resolution remains better than 350 ps.

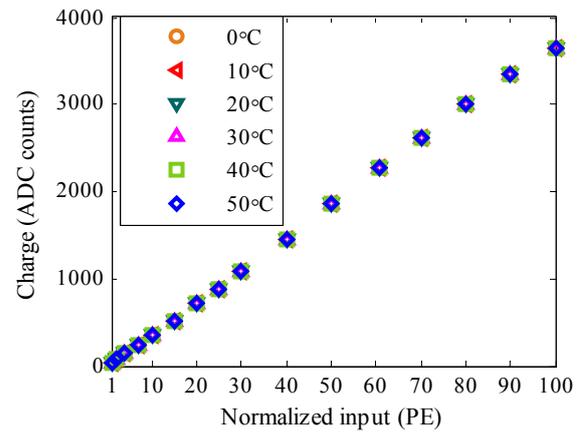

Fig. 28 Temperature dependency of charge transfer curve

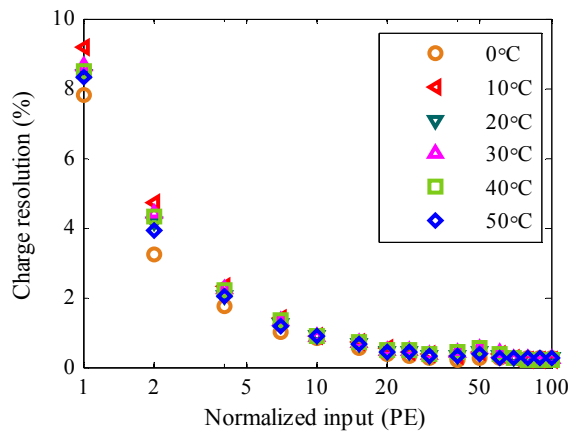

Fig. 29 Temperature dependency of charge resolution

Fig. 28 and Fig. 29 show the ambient temperature dependency of charge measurement. The charge measurement results vary within ±5% and the charge resolution remains better than 10%. Both time and charge measurement performance maintain a good consistency.

## 5. Conclusions

This paper presents the design and testing of an analogue front-end ASIC prototype for PMT signal readout. This ASIC aims to fulfill precise time and charge measurement requirement in LHAASO WCDA. The circuits are optimized to be low noise, high dynamic swing, and temperature insensitive. A series of tests have been conducted to evaluate its performance. Test results indicate that the time resolution is better than 350 ps in the full dynamic range and charge resolution is better than 10% @ 1 P.E. and better than 1% for input signals larger than 300 P.E. Meanwhile, this ASIC has been proven to have high channel-to-channel isolation and weak ambient temperature dependency.